\documentclass[aps,prl,groupedaddress,showpacs,showkeys,onecolumn]{revtex4-1}

\usepackage{graphicx} \usepackage{url} \usepackage{amsmath}
\usepackage{color}
\begin{document}

\title{Online optimization of storage ring nonlinear beam dynamics}

\author{Xiaobiao Huang }
\email[]{xiahuang@slac.stanford.edu} 
\author{James Safranek}
\affiliation{SLAC National Accelerator Laboratory, Menlo Park, CA 94025}

\date{\today}

\begin{abstract}
We propose to optimize the nonlinear beam dynamics 
of existing and future storage rings with direct online optimization techniques. 
This approach may have crucial importance for the implementation of diffraction limited 
storage rings. 
In this paper considerations and algorithms for the online optimization approach are discussed. 
We have applied this approach to experimentally improve the dynamic aperture of the SPEAR3 
storage ring with the robust conjugate direction search method and the particle swarm optimization method. 
The dynamic aperture was improved by more than 5~mm within a short period of time. 
Experimental setup and results are presented. 
\end{abstract}
 
\pacs{41.85.-p, 29.20.db, 29.27.Bd, 29.85.Fj}

\maketitle

\section{\label{intro}Introduction}
In a storage ring accelerator, sextupole magnets are used to cancel the energy dependence of 
the focusing force of quadrupole magnets (i.e., chromaticity correction). 
The magnetic forces acting on the beam by the sextupoles are nonlinear with respect to the transverse 
position of the particles. These periodic nonlinear forces can drive the motion of particles 
with large oscillation amplitude to nonlinear resonances or even chaotic regime to cause beam loss. 
Reducing the impact of the nonlinear forces of the sextupoles in order to gain large dynamic 
aperture and momentum aperture is crucial for the lattice design of a storage ring,   
and it has become more challenging  
in recent years as new rings push for lower emittances, which typically require stronger focusing and 
more lattice periods and in turn more and stronger sextupole magnets. 
For the new generation of storage ring light sources, that are based on multi-bend achromat lattice 
structure and whose emittances reach diffraction limit levels, the nonlinear dynamics challenge 
is especially severe. 
Existing storage rings have benefited greatly from robust, beam-based linear optics correction~\cite{SafranekLOCO}.
Future ultra-low emittance storage rings may require similar corrections for nonlinear optics. 

To attain large dynamic aperture, sextupoles in a storage ring are usually placed with carefully chosen 
betatron phase advances between them such that the nonlinear 
kicks on the beam by these magnets partially or fully cancel. 
Additional sextupole families (harmonic sextupoles) are often used in newer rings to cancel the 
driving terms of major lower-order resonances~\cite{BengtssonSLS}. 
For both schemes cancellation of the nonlinear kicks relies on accurate betatron phase advance control 
between the sextupole magnets. In reality, however, even after successful linear optics correction, 
it is still common to have  residual phase errors on the order of, e.g., $\sigma_\phi \sim 0.005-0.01$~rad.  
In the case of a large ring with many sextupole magnets, the incomplete cancellation could add up to give large 
errors to the resonance 
driving terms (RDTs), which may cause the nonlinear beam dynamics of the real ring to deviate significantly 
from the design model. 

Another source of errors is the differences of the nonlinear elements in the design model and the real ring. 
These include strength errors of the sextupole magnets, systematic (if not already included in the model) and  
random multipole errors of magnets and effects of insertion devices. 
Differences are also introduced by simplifications that are made in the lattice model, such as the use of the 
"hard-edge" magnet model. 
Another effect that is commonly ignored in lattice modeling is the cross-talk or interference 
between the magnetic fields of adjacent magnets. 
Differences between lattice model and real machine have been observed at the SPEAR3 storage ring~\cite{SPEARPAC09}. 

In recent years advanced multi-objective optimization methods have been widely used for the optimization 
of nonlinear beam dynamics in storage ring lattice design~\cite{BorlandGA, YangGANSLS2, HuangPSO}, 
which in simulation can significantly improve the dynamic and momentum apertures. 
However, given the potential differences between the real machine and the lattice model, it is likely that 
the optimal condition as determined by the model optimization will not give the expected 
result in the machine. In such a case, experimental correction of the nonlinear optics of the machine may be used 
to bring the machine to the optimal condition. 

Previous beam-based nonlinear dynamics optimization work includes measurement and correction of 
certain nonlinear  
dynamics characteristics such as amplitude-dependent detuning~\cite{Bartolini2011} and RDTs~\cite{Bartolini2008,Franchi}. 
However, these efforts have not yielded definitive evidence that indicates such correction improved the nonlinear dynamics 
performance measures such as the dynamic  and momentum apertures. 
Simulations  show that  minimizing driving terms does not always lead to bigger 
dynamic aperture~\cite{YangGANSLS2}. 
Therefore, it is possible that correction of the indirect 
measures will not fully recover the predicted performance. 

In this paper, we propose to use beam based measurement to directly optimize the nonlinear dynamics performance 
with systematic online optimization algorithms. In the following we discuss  considerations of the online nonlinear
dynamics optimization approach and present experimental results of the application of this approach to the SPEAR3 storage ring.

\section{Considerations of online optimization of nonlinear beam dynamics}\label{secConsider}
The nonlinear dynamics performance of a storage ring is ultimately characterized by the dynamic aperture
and the momentum aperture, which affect  two important operational performance measures, the injection efficiency and 
the Touschek lifetime, respectively. 
In fact, the purpose of nonlinear dynamics optimization for a storage ring is to obtain good injection efficiency and 
long Touschek lifetime. 
In addition, measuring dynamic aperture and momentum aperture is usually more time consuming than directly measuring 
injection efficiency and beam lifetime. 
Therefore the objectives of experimental optimization of ring performance would be injection efficiency and 
beam lifetime (with lifetime measured under certain fixed, Touschek lifetime dominated machine conditions). 
Since the evaluation of beam lifetime requires steady beam condition which is not compatible with frequent injection,  
it is difficult to simultaneously optimize the two objectives for a given lattice. 
One approach may be to optimize one objective  at a time and iterate if necessary. 

The optimization variables (i.e. knobs) can be the strengths of sextupole magnets. 
Since sextupole magnets located in dispersive regions affect the chromaticities, the sextupole knobs should be combined 
to form independent variables that do not change chromaticities. 
One way to form such variables is described in the next section. 
For large rings with an individual power supply for each sextupole magnet, the number of free variables may be reduced by
grouping  the power supplies accordingly to symmetry or periodicity considerations 
to form individual knobs. 
It is also possible to use combinations of quadrupole magnets as optimization variables,  
if these knobs are set up to meet the linear optics requirements of the storage ring.  

Efficiency and robustness are two key requirements for online optimization algorithms. 
High efficiency means the algorithms need to find the minimum with a small number of evaluations of the objective 
function(s). Robustness requires the algorithms to be effective under the influence of random noise and occasional 
outliers in the measurement of the objective function(s).
Ref.~\cite{RCDS} proposed the robust conjugate direction search (RCDS) method for online optimization 
and demonstrated through simulation and experiments that it is both efficient and robust. 
The RCDS algorithm is a single-objective method and would stop at a local minimum. 
However, it is usually a good choice for online nonlinear dynamics optimization, because the 
injection efficiency and lifetime objectives normally need to be optimized separately and in most cases 
the global minimum is not far from the starting point, which is typically set according to an optimized model. 
Convergence is fast in the vicinity of the minimum for conjugate direction search methods such as RCDS. 

The RCDS may also be used in the case when the starting point is far from the global minimum (e.g., for the linear 
coupling correction problem as demonstrated in Ref.~\cite{RCDS}). 
However, if a more thorough search of the parameter space is desired in order to gain more assurance that the global minimum 
is reached, stochastic optimization algorithms such as genetic algorithms~\cite{Bazarov2005,BorlandGA, YangGANSLS2} and 
particle swarm optimization (PSO) algorithms~\cite{BorlandPSO,Pang2014124,HuangPSO} 
can be used. Studies have indicated the particle swarm method is a few times more efficient than 
a widely used genetic algorithm because solutions evaluated by the former are considerably more diverse than that of the 
latter. In addition, Ref.~\cite{RCDS} showed that, for genetic algorithms, solutions biased by random errors tend to 
enter the next generation and therefore hamper the convergence to the true global minimum. 
For these reasons the particle swarm method may be the preferred stochastic optimization algorithm for online 
applications. 

In the following we will discuss the online nonlinear dynamics optimization for the SPEAR3 storage ring with 
the RCDS and PSO methods, which 
may serve as an illustration of the above considerations.

\section{Online optimization of dynamic aperture for SPEAR3 }\label{secExperiment}
SPEAR3 is a third generation storage ring light source. 
It has a total of 72 sextupoles, 36 focusing sextupoles (SF) and 36 defocusing sextupoles (SD), 
with two SFs and two SDs in each of its 18 cells (14 standard cells and 4 matching cells).
The 72 sextupoles used to be powered by 4 power supplies 
for chromaticity correction. 
During the 2014 summer shutdown, 6 additional sextupole power supplies were added, resulting in a total of 10 sextupole 
families. Each family consists of several sextupoles of the same type (SF or SD). 
These extra sextupole power supplies enabled the online nonlinear dynamics optimization to be discussed below. 

The response matrix of chromaticities  with respect to the 10 sextupole families, ${\mathbf R}_\xi$, is 
calculated using the lattice model. To keep chromaticities fixed during optimization, it is sufficient and necessary that 
any changes to the sextupoles be in the null space of ${\mathbf R}_\xi$, which is a 2 by 10 matrix. 
With singular value decomposition (SVD), ${\mathbf R}_\xi={\mathbf U }{\mathbf S }{\mathbf V}^T$, the null 
space is simply the sub-space spanned by the 8 vectors in ${\mathbf V}$ which correspond to zero singular values. 
These 8 vectors represent a basis of the null space of the chromaticity response matrix, so they were 
used as the free knobs in our nonlinear dynamics optimization. 

When using the RCDS method, it is recommended to find an initial conjugate direction set in order to improve searching 
efficiency~\cite{RCDS}. 
Such a direction set could be obtained with the Jacobian matrix of some selected nonlinear RDTs or 
the distortion functions as proposed by Collins~\cite{CollinsDF} with respect to the 8 free knobs. 
But since it is unclear if the direction set will be effective, and for simplicity, 
we simply used the 8 unit vectors as the initial directions. 

The objective function is the injection efficiency. It is evaluated as the increase of stored beam current 
divided by the average intensity of the injector beam for an interval of 10 seconds. 
Noise in the objective function comes mainly from the uncertainty of the injected beam intensity measurement. 
The noise sigma of injection efficiency is about 3\%. 

\begin{figure}[hbt]
  \centering
  \includegraphics[width=2.5in]{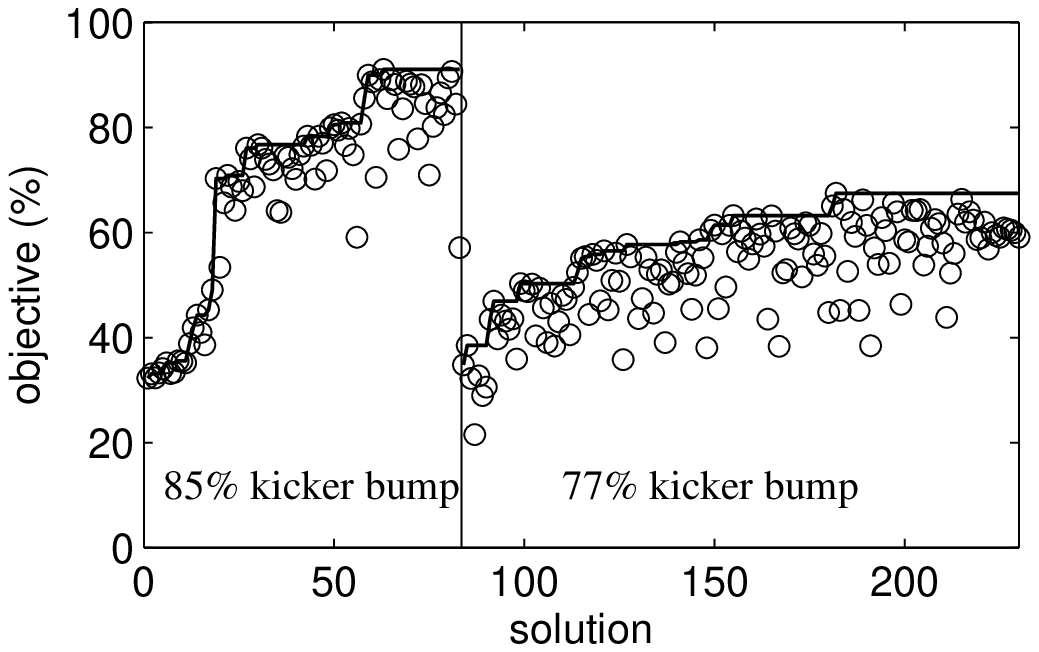}
  \includegraphics[width=2.5in]{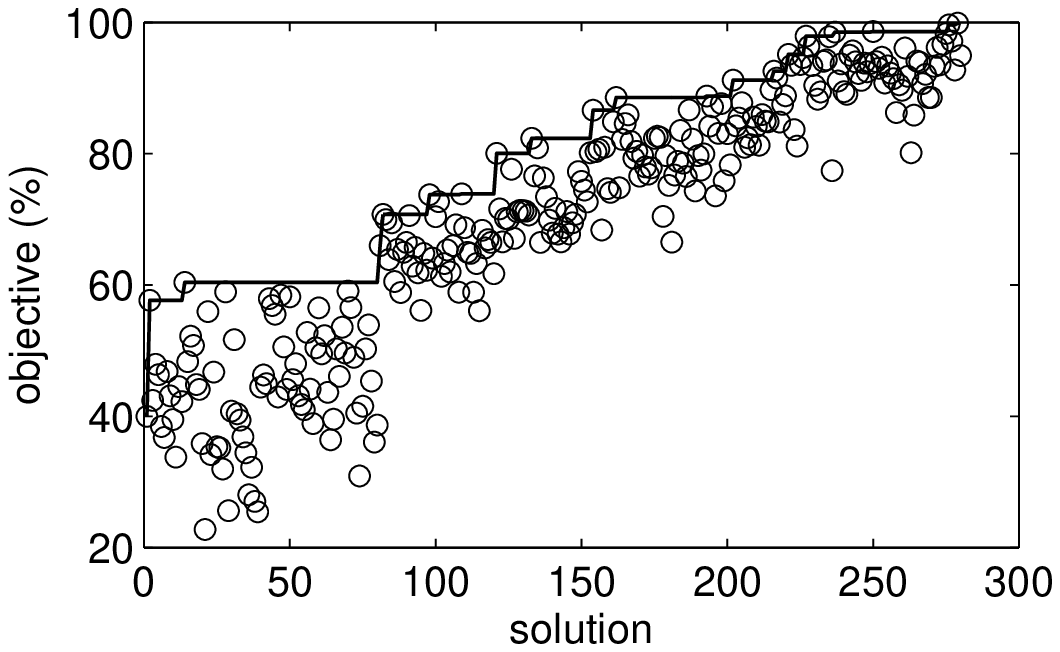}
  \caption{\label{figHistObj} History of objective function values of all evaluated solutions during the RCDS optimization (left) 
  and the PSO optimization (right).  }
\end{figure}

We have applied online dynamic aperture optimization for the SPEAR3 operational lattice with both the RCDS 
and the PSO algorithms.  
Since this lattice normally has good injection efficiency, we intentionally scaled down the injection kicker bump,  
which effectively increased the dynamic aperture requirement for 
capturing the injected beam. 
FIG.~\ref{figHistObj} left plot shows the history of the objective function of all evaluated solutions during an   
RCDS optimization. In this experiment, the kicker bump was first reduced to 85\% 
of the original full bump,
and then was further reduced to 77\%
as injection efficiency was improved. 
This RCDS run was terminated after three iterations of the direction set scans were completed. 
The PSO method was applied to the same problem after a sizable improvement of dynamic aperture was made with a 
preliminary RCDS run. The population of 
solutions for the PSO run was 40. The initial population was centered around the solution (which was included) found by the preliminary RCDS 
run. Deviation of each 
variable value of the initial solutions was drawn from 
a random uniform distribution, whose size is $\pm$ 10\%
of its range. The kicker bump was reduced so that the initial injection efficiency was about 40\%. 
The algorithm was run for 7 generations and was stopped when the injection efficiency reached 100\%. 
The history of the objective function of the PSO optimization was shown in FIG.~\ref{figHistObj} right plot. 

\begin{figure}[hbt]
  \centering
  \includegraphics[width=4.5in]{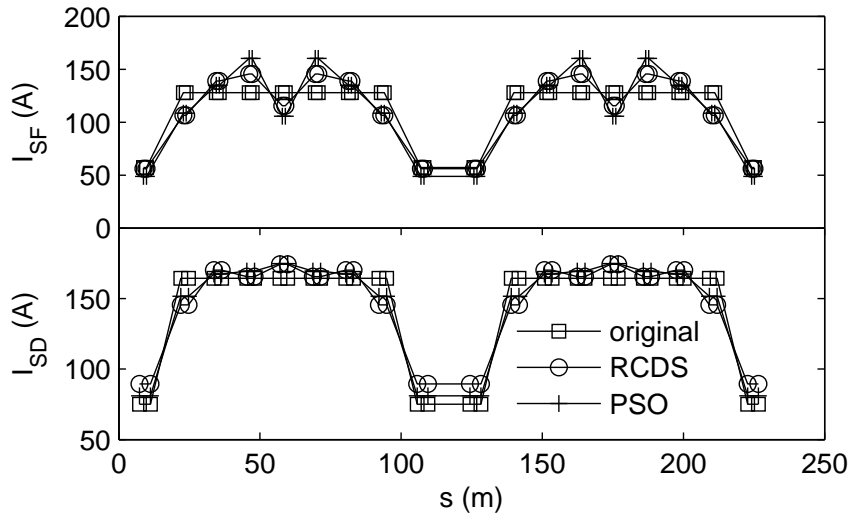}
  \caption{\label{figCmpSext} Comparison of sextupole settings for the original and optimized solutions.}
\end{figure}

The sextupole settings of the RCDS and PSO solutions are compared to the original values in 
FIG.~\ref{figCmpSext}. The sextupole strengths in the standard cells for the optimized solutions 
are changed substantially. The solutions found by PSO and RCDS are very similar. 
Chromaticities were measured and it was found that the horizontal and vertical chromaticities of all   
optimized solutions were equal to the original values of $\xi_x=\xi_y=+3$. 

\begin{figure}[hbt]
  \centering
  \includegraphics[width=2.5in]{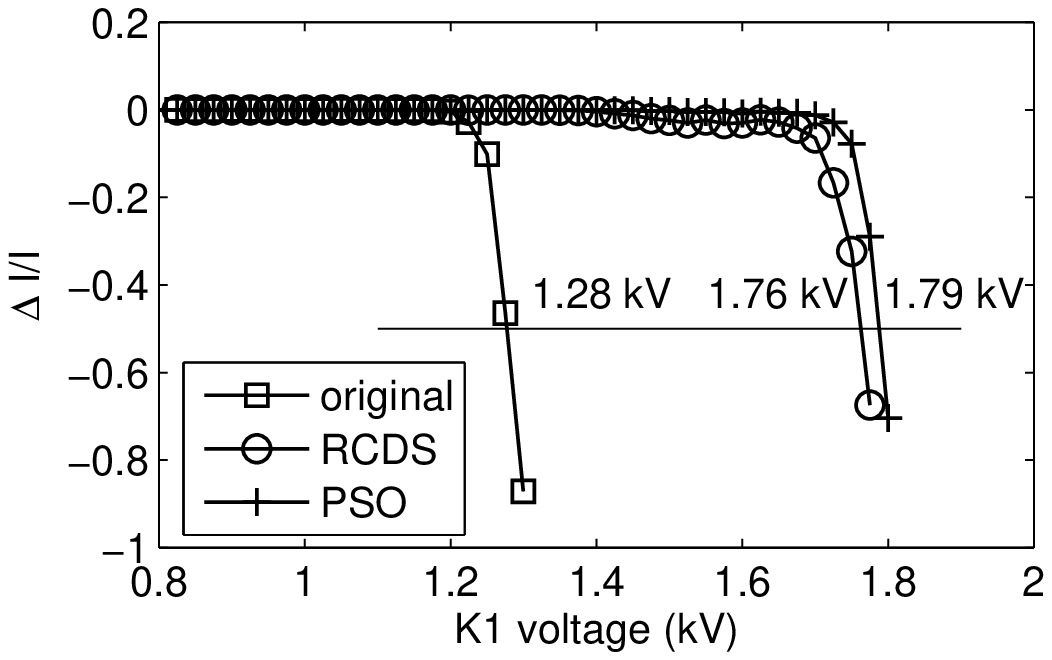}
  \includegraphics[width=2.5in]{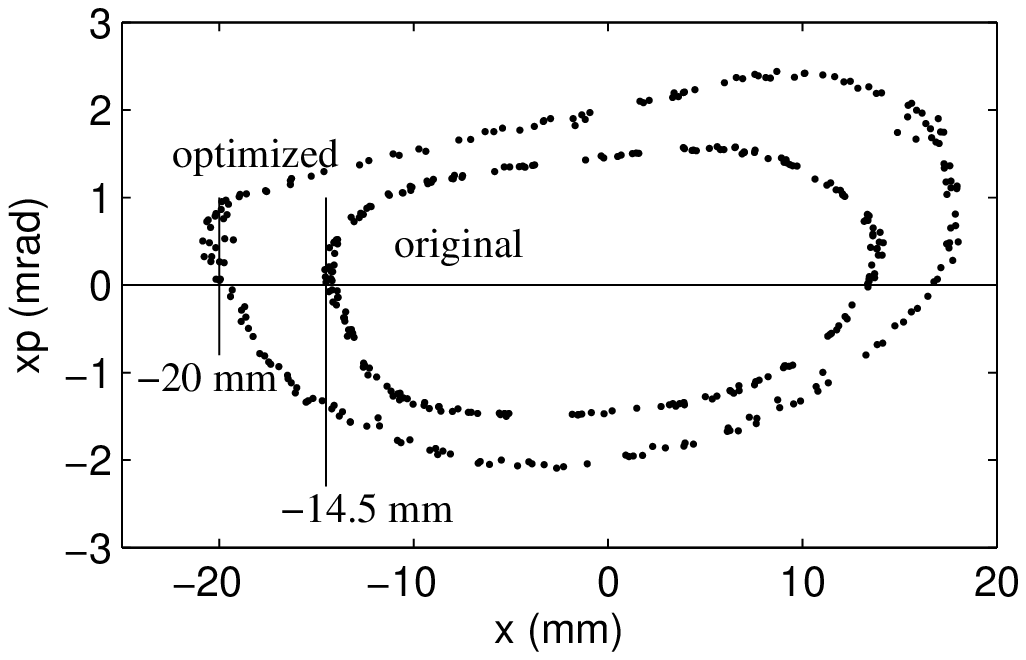}
  \caption{\label{figCmpDA} Comparison of measured dynamic aperture for the original and optimized solutions. 
   }
\end{figure}

Dynamic aperture measurement was done to characterize the optimized solutions. In this measurement a short 
train of bunches of stored beam was kicked by one of the injection kickers with increasing kick strength until 
all beam was lost. The kicker was fired 10 times at each kick strength. 
FIG.~\ref{figCmpDA} left plot shows the fractional beam current change vs. kicker voltage, where 
an arbitrary line at 50\% 
of fractional loss indicates the kick strengths that drive the beam to the edge of the dynamic aperture 
for the three cases. The  kick angles are used in particle tracking with the corresponding lattice models 
for 200 turns. The resulted phase space plots for the two optimized solutions are too close to discern. 
In FIG.~\ref{figCmpDA} right plot we only show a comparison of the PSO solution ("optimized") and the 
original sextupole setting.
If the half size of the stored beam ($2\sigma_x=0.6$~mm) is included, the dynamic aperture for the 
original and optimized solutions are 15.1 and 20.6~mm, respectively. 

\begin{figure}[hbt]
  \centering
  \includegraphics[width=4.0in]{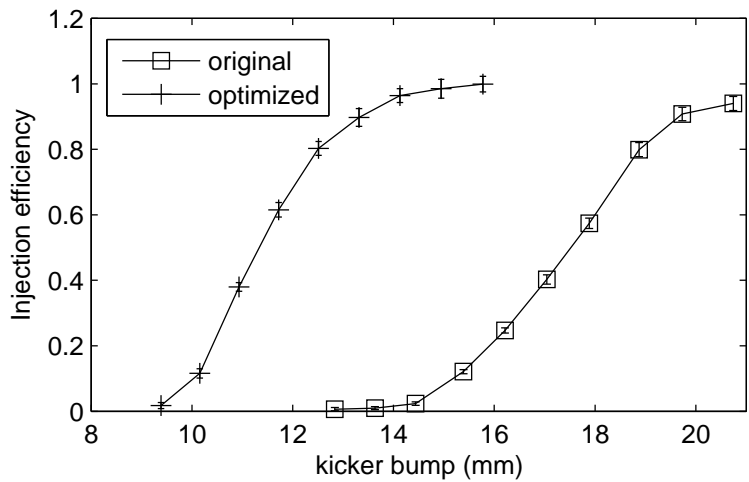}
  \caption{\label{figCmpInjEffvsKbump} Injection efficiency vs. kicker bump for the original and optimized solutions.}
\end{figure}

The dynamic aperture gain for the optimized solutions was further verified with the measurement of injection efficiency 
vs. the injection kicker bump. The strengths of all three injection kickers were scaled  
to vary the size of the kicker bump and the injection efficiency was checked at 
each step. The results for the original and optimized sextupole settings are shown in FIG.~\ref{figCmpInjEffvsKbump}. 
The kicker bumps for both solutions were closed (i.e., there was no residual oscillation 
for the stored beam.) for the measurements. 
The measurements confirmed that the dynamic aperture was increased by more than 5~mm after the online optimization.  

Beam lifetime under operation conditions (500 mA in 280 bunches, with RF voltage of 2.85~MV)  was found to be about 
6.8 hrs for the optimized solutions, compared to 7.8 hrs for the original sextupole setting. 
To check for possible reduction of momentum aperture for the optimized solutions, we measured the beam lifetime under 
the Touschek lifetime dominated regime while reducing the RF voltage.
It was found that for both the RCDS and PSO solutions the lifetime went down  monotonically with a decreasing RF voltage 
when it was below 3.1~MV, which indicates 
that the momentum aperture is determined by the bucket height rather than a reduced off-momentum 
dynamic aperture~\cite{SteierPRE65}. Therefore, the reduced lifetime at 500 mA for the optimized solutions is likely due to  changes of 
vertical emittance that resulted from the changes of sextupole setting.  

\begin{figure}[hbt]
  \centering
  \includegraphics[width=4.0in]{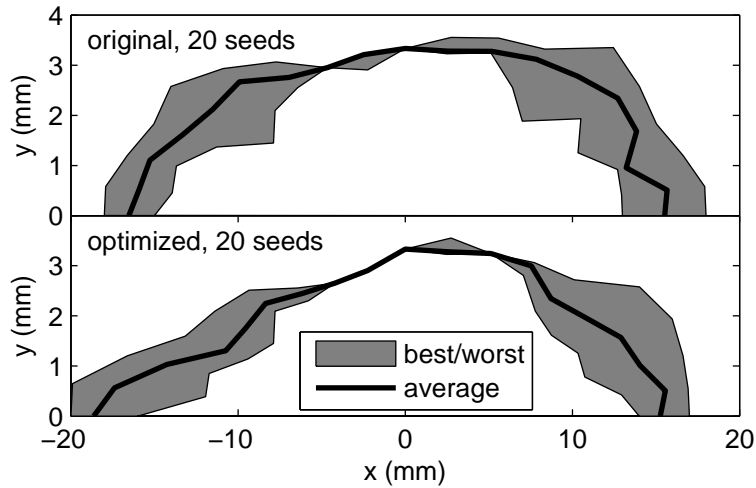}
  \caption{\label{figCmpDATrack} Dynamic aperture obtained by particle tracking for the original 
  (top) and optimized (bottom) solutions.}
\end{figure}

The dynamic apertures for the original and optimized sextupole settings have been evaluated with particle tracking simulation. 
The simulation includes realistic vertical physical aperture, effects of insertion devices and systematic magnetic multipole errors.
Random linear and nonlinear errors are also added with 20 random seeds. The linear optics error was corrected to 1\% 
rms beta beating 
and 0.2\% 
linear coupling, close to realistic values. Results of simulated dynamic aperture for the original and the optimized solutions 
are shown in FIG.~\ref{figCmpDATrack}. 
Interestingly, the dynamic aperture determined by tracking for the optimized  
solution indeed has a substantial gain on the horizontal mid-plane. But there is a dent on the vertical direction 
near $x=-10$~mm. 
If this dent is real for the actual machine, it indicates the experimentally optimized solution found a better balance 
between the horizontal and vertical dynamic aperture requirements. 
Optimizing the dynamic aperture and momentum aperture using the model in simulation for a given random error seed 
with the multi-objective particle swarm 
method yielded solutions with dynamic aperture that extends to $-20$~mm with no vertical dent. 
Therefore it is also possible that the optimized real machine has no such a dent in the dynamic aperture and 
the discrepancy may come from the differences between the lattice model and the machine. 

\section{Conclusion}
We propose to optimize the nonlinear dynamics of storage rings online directly with advanced optimization algorithms.
Because of the intrinsic differences between real machines and their corresponding lattice models, ideal performance 
of optimized models may not be directly achieved on the real machine. The online approach could bridge the gap between 
the optimized model and the actual machine to achieve the optimal performance. 

We have demonstrated the power of the online methods with dynamic aperture optimization for the SPEAR3 storage ring. 
In the experiments the dynamic aperture of SPEAR3 was improved from 15.1 mm to 20.6 mm, while the momentum aperture  
was not negatively affected. 

\begin{acknowledgments}
The study is supported by DOE Contract No. DE-AC02-76SF00515. 
\end{acknowledgments}

\bibliography{da_ref}

\end{document}